# Magnetic tunnel junctions of twisted $La_{2/3}Sr_{1/3}MnO_3$ freestanding bilayers

Kota Kanda[1], Shiho Irie[1], Kodai Niitsu[2], Masayuki Ochi[1,3], Takumi Yamazaki[4], Kouta Kondou[5], Kohei Ueda[1,5], Takeshi Seki[4,6,7], Jobu Matsuno[1,5], and Junichi Shiogai[1,5,†]

[1]*Department of Physics, The University of Osaka, Toyonaka, Osaka 560-0043, Japan*

[2]*Center for Basic Research on Materials, National Institute for Materials Science (NIMS), Tsukuba, Ibaraki 305-0047, Japan*

[3]*Forefront Research Center, The University of Osaka, Toyonaka, Osaka 560-0043, Japan*

[4]*Institute for Materials Research, Tohoku University, Sendai, Miyagi 980-8577, Japan*

[5]*Division of Spintronics Research Network, Institute for Open and Transdisciplinary Research Initiatives, The University of Osaka, Suita, Osaka 565-0871, Japan*

[6]*Center for Science and Innovation in Spintronics, Tohoku University, Sendai, Miyagi 980-8577, Japan*

[7]*International Center for Synchrotron Radiation Innovation Smart, Tohoku University, Sendai, Miyagi 980-8577, Japan*

[†]Corresponding author: junichi.shiogai.sci@osaka-u.ac.jp

**Transition metal oxides (TMOs) host a variety of emergent spintronic or multiferroic functionalities owing to the intricate interplay between lattice, orbital, charge, and spin degrees of freedom. Recent advances in epitaxial lift-off and dry transfer techniques using water-soluble sacrificial layers enables stacking and twisting of freestanding membranes of such strongly correlated oxides, providing even richer physical properties and functionalities. To date, the twisted junctions of TMOs have been mainly investigated in insulators, focusing on their structural and dielectric properties. Here, we demonstrate that the twisted bilayer membranes of $La_{2/3}Sr_{1/3}MnO_3$, a prototypical half-metallic ferromagnetic metal TMO, offer a magnetic tunnel junction (MTJ) with their van der Waals (vdW) gap acting as a tunnel barrier. The twisted junctions of $La_{2/3}Sr_{1/3}MnO_3$ exhibit tunnelling magnetoresistance (TMR) effect with respect to relative orientation of magnetizations of top and bottom membranes. In addition, high-field magnetoresistance shows uniaxial anisotropy, which can be assigned to tunnelling anisotropic magnetoresistance (TAMR) effect reflecting rhombohedral distortion of fully-relaxed $La_{2/3}Sr_{1/3}MnO_3$. The demonstration of TMR and TAMR effects suggests that spin and in-plane momentum degrees of freedom of 3*d* electrons are conserved across the twisted junctions of TMOs, opening a new avenue for realizing**

**strongly correlated twistronics utilizing electron coherence through tunnelling processes.**

The vdW junctions of atomically thin two-dimensional (2D) materials, including graphene, hexagonal boron nitride, and transition metal chalcogenides, have been a versatile platform for exploring emergent interfacial physical properties and device functionalities[1,2]. Particularly in twisted vdW junctions, precise adjustment of the twist angle between the 2D materials introduces a moiré superlattice into the junction, creating strongly correlated interfaces associated with flat-band electronic states[3,4]. To date, exotic interfacial physical properties such as resonant tunnelling[5], unconventional superconductivity[6], ferromagnetism[7], ferroelectricity[8], as well as quantum anomalous Hall states[9] have been reported in twisted junctions of 2D materials, opening a new field of twistronics. Transition metal oxides (TMOs) host rich strongly correlated physical properties including half-metallic ferromagnetism, ferroelectricity, metal-insulator transition as well as high-temperature superconductivity owing to intricate couplings between lattice, orbital, charge, and spin degrees of freedom of electrons[10,11]. Analogous to the 2D materials, engineering of lateral twist angles $\theta$ in the vdW junctions of 2D crystals of such strongly correlated TMOs (Fig. **1a**) would lead to even richer interfacial physical properties and functionalities absent in conventional epitaxial thin-film interfaces[12,13].

Recent advances in epitaxial lift-off technique using water-soluble sacrificial layers[14,15,16] have realized such twisted junctions of freestanding membranes of various TMOs[17,18,19,20]. So far, the intriguing interfacial structural properties of the twisted junctions of insulating TMOs have been reported, primarily focusing on emergence of polar vortex structures in moiré superlattices of ferroelectric $BaTiO_3$ (Ref. 21) and paraelectric $SrTiO_3$ (Refs. 22,23,24). In contrast, observation of electrical transport across the vdW junctions of conducting TMOs poses a significant challenge because the interfaces synthesized by the epitaxial lift-off technique suffer from process-borne contamination[19,25,26].

In contrast to the naturally cleaved vdW surfaces of 2D materials in mechanical exfoliation, the surfaces of freestanding TMO membranes with 3D ionic bonding are inevitably exposed to sticky polymer supporters used for pickup and transfer during the epitaxial lift-off process or to adsorbed species. These process-borne contaminations hinder formation of the sufficiently clean interfaces required for conservation of electron coherence across the vdW junctions. In our experiments, we established the process to synthesize the clean twisted junctions of ferromagnetic $La_{2/3}Sr_{1/3}MnO_3$ (Fig. **1b**), a prototypical half-metallic TMO, by applying the aluminum oxide ($AlO_x$) cap layer to protect their surface from the process-borne contaminations and removing it prior to the stamping the second layer to expose the clean surface of $La_{2/3}Sr_{1/3}MnO_3$. With this process, we realized twisted MTJs of $La_{2/3}Sr_{1/3}MnO_3$ exhibiting twist angle $\theta$ dependent TMR and TAMR effects, which is a clear signature of conservation of electron spin and in-plane momentum degrees of freedom across the junctions (Fig. **1c**). It turns out that the vdW gap of the twisted $La_{2/3}Sr_{1/3}MnO_3$ membranes, formed without intentional insertion of an insulator, serves as a tunnelling barrier where the quantum coherence of electrons persists in the tunnelling process.

**Fabrication of twisted junctions of $La_{2/3}Sr_{1/3}MnO_3$ membranes.**

The twisted junctions were fabricated by exfoliating and transferring two freestanding 80-nm-thick $La_{2/3}Sr_{1/3}MnO_3$ membranes from $AlO_x$-capped $La_{2/3}Sr_{1/3}MnO_3/Sr_3Al_2O_6$ epitaxial thin-film heterostructure on $SrTiO_3(001)$ substrate using epitaxial lift-off technique as illustrated in Fig. **1d**. After transferring an $AlO_x/La_{2/3}Sr_{1/3}MnO_3$ freestanding membrane using polydimethylsiloxane (PDMS) on Si substrate, the amorphous $AlO_x$ cap layer was removed by Ar ion milling to expose the clean surface of $La_{2/3}Sr_{1/3}MnO_3$ as a bottom layer. Then, the second membrane was stamped as a top layer, where the twist angle $\theta$ was defined by the relative angle of [100] crystallographic direction of the top layer with respect to that of the bottom layer anticlockwise. Details of sample fabrication are given in Materials and Methods section and Supplementary Section **1**. Electrical transport properties of the twisted junctions were evaluated by two-terminal resistance ($R_J$) and three-terminal resistances ($R_{3T}^{top}$ and $R_{3T}^{btm}$). The $R_J$ corresponds to the junction resistance across the Pt/$La_{2/3}Sr_{1/3}MnO_3$/$La_{2/3}Sr_{1/3}MnO_3$/Pt, while $R_{3T}^{top}$ and $R_{3T}^{btm}$ measure the resistances across Pt/$La_{2/3}Sr_{1/3}MnO_3$ in top and bottom membranes, respectively (see also Supplementary Figs. **S2a** and **S2b**).

**Structural and electrical transport characterizations of the twisted junctions.**

Figure **2a** shows a top-view photograph of an exfoliated membrane on PDMS supporter after lift-off in water. Many cracks are formed along <110> (45° tilted from [100]/[010]) or <100> directions (shown in Supplementary Fig. **S1c**), corresponding to the twin boundaries owing to the rhombohedral distortion along <111> of freestanding $La_{2/3}Sr_{1/3}MnO_3$ (Ref. 27). It verifies that each exfoliated flake consists of a single rhombohedral domain of $La_{2/3}Sr_{1/3}MnO_3$. Figure **2b** shows a scanning electron micrograph of the final sample geometry of the typical twisted junction. For microscopic structural analysis around the twisted junction, cross-sectional bright field (BF) transmission electron microscope (TEM) and high-angle annular dark field (HAADF) scanning transmission electron microscope (STEM) images were taken for a typical 45°-

junction. The wide-area BF-TEM image shows that the top and bottom $La_{2/3}Sr_{1/3}MnO_3$ membranes are separated by a uniform gap without residual contaminants as shown in Fig. **2c**. From the enlarged HAADF-STEM image (Fig. **2d**), the thickness of the formed gap is estimated to be approximately 3 nm. This cross-sectional structural analysis verifies that the vdW junction is formed by the $La_{2/3}Sr_{1/3}MnO_3$ bilayer. The electron diffraction patterns, acquired separately from the top and bottom membranes (insets in Fig. **2d**) reveal that the misalignment of $\theta$ is below 7°.

Figure **2e** shows temperature ($T$) dependences of $R_J$ (red solid line), $R_{3T}^{top}$ (green), and $R_{3T}^{btm}$ (blue) for a 45°-junction. The $R_J(T)$ exhibits an insulating behaviour, which is distinct from metallic properties observed in $R_{3T}^{top}(T)$ and $R_{3T}^{btm}(T)$ below the kink at $T \sim 180$ K and 250 K (lower triangles), respectively. The kink temperatures in $R_{3T}^{top}(T)$ and $R_{3T}^{btm}(T)$ around the metal-insulator transition are ascribed to ferromagnetic transition temperature $T_C$ of the $La_{2/3}Sr_{1/3}MnO_3$ below the Pt electrode. Despite that the bulk $T_C$ of the freestanding membrane of $La_{2/3}Sr_{1/3}MnO_3$ is well above room temperature (see Supplementary Section **3**), that of the surface is reduced by the presence of a few nm dead layer as reported previously[28].

The tunnelling barrier properties were investigated from a current-voltage (*I*-*V*) characteristic across the junction. Figure **2f** shows a clear non-linear *I*-*V* curve obtained for the 45°-junction measured at $T$ = 4.2 K. The $I$-$V$ characteristic was analyzed by Simmons' formula assuming the tunnel conductance[29] (a black dashed line in Fig. **2f**), yielding that effective barrier height $E_{eff}$ is 0.27 eV and thickness $d_{eff}$ is 3.5 nm. Similar tunnelling-like $I$-$V$ curves were observed for other twisted junctions (see Supplementary Section **4**). These insulating $R_J(T)$ and non-linear $I$-$V$ characteristics across the twisted vdW junctions of two ferromagnetic $La_{2/3}Sr_{1/3}MnO_3$ membranes are direct evidence of formation of the tunnel junction. Figure **2g** shows temperature dependence of a product of $R_J$ and junction area $A$ for junctions with different $\theta$. The overall value of $R_JA$ product increases with increasing $\theta$ although $E_{eff}$ and $d_{eff}$ are comparable. It can be speculated that the net area where the gap between the top and bottom layers is narrowest is smaller for larger $\theta$ due to interface reconstruction by annealing process[30,31].

**Observation of the tunnelling magnetoresistance effect.**

When electron spins are conserved in the tunnelling process, the junction resistance should depend on the relative orientation of magnetization vectors (**M**) of top and bottom ferromagnetic electrodes, resulting in the TMR effect[32,33]. Figure **3a** shows typical MR curves of $R_J$ at $T$ = 2 K for $\theta$ ~ 0° and 45°-junctions, when magnetic field $B$ was applied along the [100] direction of the bottom $La_{2/3}Sr_{1/3}MnO_3$ membranes (see the inset) and the bias voltage $V_b = R_J I \sim 40$ mV was applied. A clear positive spin-valve-like MR effect at around $B \sim 20$ mT is commonly observed for both junctions (vertical red and black shaded areas), which is linked to the reversal of **M** in the bottom membrane as confirmed by the butterfly-shaped MR in $R_{3T}^{btm}(B)$ (filled triangles in the bottom data in Fig. **3a**). The closure of the positive MR occurs at higher $B$, because coercive field of the top membrane, which is not exposed to milling process, is higher than that for milled membrane (see Supplementary Fig. **S3a**). The fact that the spin-valve effect being dependent on the relative angle of **M** for top and bottom ferromagnets is a hallmark of the observation of TMR effect in the twisted junction of the $La_{2/3}Sr_{1/3}MnO_3$ membranes. Note that possible superposition of the anisotropic magnetoresistance (AMR) in a single $La_{2/3}Sr_{1/3}MnO_3$ membrane can be excluded because the lateral resistance of the single membrane is three orders of magnitude smaller than $R_J$.

Figure **3b** shows evolution of the spin-valve effect with increasing bias voltage $V_b$ for the 45°-junction. Overall, the MR ratio (indicated by a vertical arrow) monotonically decreases with increasing $V_b$, being consistent with the previously reported TMR effect on the $La_{2/3}Sr_{1/3}MnO_3/SrTiO_3/La_{2/3}Sr_{1/3}MnO_3$ epitaxial MTJ (Ref. 34). In the two-flow model, the TMR ratio is given as[33]

$$\text{TMR ratio (\%)} = \frac{R_{AP} - R_P}{R_P} \times 100 = \frac{2P^2}{1 - P^2} \times 100. \qquad \text{Eq. (1)}$$

The $R_P$ and $R_{AP}$ represent resistances in the parallel and antiparallel configurations, respectively. We assume the same spin polarization $P$ for the top and bottom ferromagnetic electrodes. Based on Eq. (1), the observed MR ratio reaching as high as 23% at low $V_b$ corresponds to $P$ of about 32%. The extracted TMR ratio is one order of magnitude higher than that of twisted homojunctions of 2D ferromagnet $Fe_3GeTe_2$ reported previously[35], reflecting formation of a clean vdW interface and high spin polarization of $La_{2/3}Sr_{1/3}MnO_3$. Figure **3c** shows the MR ratio as a function of $V_b$ for the 0° and 45°-junctions. Insensitive to $\theta$, the MR ratio monotonically decreases with increasing $V_b$. The reduction of MR ratio in a high $V_b$ regime may be regarded as a spin wave excitation induced by the tunnelling carriers as discussed in the previous report on the epitaxial MTJs of $La_{2/3}Sr_{1/3}MnO_3$ (Ref. 34).

Temperature dependence of the TMR effect was investigated for the 0°-junction by comparing $R_J$ with anti-parallel configuration, $R_{AP}(T)$, and that with parallel configuration, $R_P(T)$. The anti-parallel configuration was obtained by sweeping $B$ from +1 T to −16.5 mT and back to zero field (minor loop, blue solid line in Fig. **3d**), while the parallel configuration was obtained by sweeping $B$ from +1 T to zero field at $T$ = 2 K. Figure **3e** shows $R_{AP}(T)$ and $R_P(T)$ curves obtained by sweeping $T$ from 2 K to 300 K. With increasing $T$, the two curves merge around $T$ = 180 K, which is consistent with the kink temperature observed in $R_{3T}^{top}(T)$ in Fig. **2e**. *i.e.*, the resistance between Pt electrode and the top $La_{2/3}Sr_{1/3}MnO_3$ membrane. As the kink temperatures in $R_{3T}^{top}(T)$ and $R_{3T}^{btm}(T)$ can be assigned to $T_C$ of the top and bottom $La_{2/3}Sr_{1/3}MnO_3$ membranes, the results show that the TMR effect is observed only when both top and bottom membranes are in the ferromagnetic state.

**Tunnelling anisotropic magnetoresistance effect.**

The coherent transport of electron tunnelling across the vdW junction is further evidenced by anisotropic tunnelling conductance. In addition to the TMR effect, we observed a negative sign of the spin-valve-like MR and high-field MR for a 70°-junction as shown in Fig. **4a**. According to Eq. (1), the negative spin-valve effect cannot be merely explained by TMR effect because of the same $P$ of the top and bottom electrodes. Such negative MR has been reported in epitaxial MTJs of ferromagnetic semiconductor (Ga,Mn)As (Refs. 36,37) and single ferromagnetic electrode[38,39]. This phenomenon is known as tunnelling anisotropic magnetoresistance (TAMR) effect, where $R_J$ exhibits anisotropy depending on the orientation of **M** with respect to crystallographic direction. For the sake of simplicity, we focus on the high-field $R_J$ of the 0°-junction, where crystallographic direction of top and bottom layers are parallel and **M** of both layers are oriented along magnetic field vector **B**. Figure **4b** shows $R_J(B)$ curves of the 0°-junction when **B** is applied along in-plane angle $\phi_B$ = 74° (red) and 170° (blue) with respect to the [100] crystallographic orientations of the $La_{2/3}Sr_{1/3}MnO_3$ membranes. While both curves exhibit similar spin-valve behaviour at low $B$, which we ascribed to TMR effect, the $R_J(B)$ curves do not overlap in sufficiently large in-plane $B$. It turns out that the high-field $R_J$ consists of a contribution from tunnelling conductance depending on the orientation of **M** with respect to crystallographic direction *i.e.*, TAMR effect. Symmetry of the TAMR effect was further investigated by rotating **B** in the plane[38,40] as shown in Fig. **4c**. Here, the sufficiently strong $B$ of 1 T was applied to align **M** of both top and bottom $La_{2/3}Sr_{1/3}MnO_3$ membranes. A uniaxial TAMR effect along the $[1\bar{1}0]$ direction clearly emerges while the values of $R_J$ are almost the same for **M**//[100] and [010] directions. The value of anisotropy of 6% is comparable to that observed in (Ga,Mn)As (Ref. 40) and much larger than that in Fe/GaAs/Au heterostructure[38].

As TAMR reflects the anisotropy in the band dispersion $E(k)$ of the ferromagnetic electrode with respect to **M** orientation, such a large value of TAMR provides strong evidence for the conservation of the in-plane momentum of conducting electrons during the tunnelling process, as discussed in epitaxial MTJs (Refs. 36,39). In the case of $La_{2/3}Sr_{1/3}MnO_3$, the anisotropy of tunnelling conductance is dominated by the $d_{z^2}$ orbitals because it possesses dispersive $\sigma$-bond-like coupling along the transport direction rather than planar $d_{x^2-y^2}$ orbitals. In previous study for the epitaxial $La_{2/3}Sr_{1/3}MnO_3/SrTiO_3/La_{2/3}Sr_{1/3}MnO_3(001)$ heterojunction, the in-plane TAMR stems from the spin-orbit interaction due to the hybridization of $d_{z^2}$ orbitals and interface localized states of the $t_{2g}$ bands[41]. Actually, the epitaxial MTJ consisting of a similar manganite compound grown on $SrTiO_3$ with biaxial tensile strain exhibits in-plane biaxial TAMR effect along [110] and $[1\bar{1}0]$ directions[42]. For the freestanding $La_{2/3}Sr_{1/3}MnO_3$ membrane, we discuss the breaking biaxial anisotropy of the TAMR effect in a framework of rhombohedral distortion[43]. As shown in Fig. **4d**, the bulk $La_{2/3}Sr_{1/3}MnO_3$ exhibits the rhombohedral distortion along [111] direction (indicated as an orange arrow) associated with the rotation of neighbouring $MnO_6$ octahedra in the opposite directions[27]. On the (001) plane, all the $MnO_6$ octahedral are tilted along $[1\bar{1}0]$ as indicated by white arrows in Fig. **4e**, making the [110] and $[1\bar{1}0]$ directions are inequivalent. Under this condition, the $E(k)$ can be different between for **M** along [110] and $[1\bar{1}0]$ in the presence of spin-orbit interaction, which is qualitatively supported by density functional theory in bulk $La_{2/3}Sr_{1/3}MnO_3$ around Fermi energy (see Supplementary Section **5**). Considering the observation of uniaxial TAMR effect, we conclude that the tunnelling conductance in 2D Brillouin zone reflects uniaxially distorted crystal symmetry of $La_{2/3}Sr_{1/3}MnO_3$ through the in-plane momentum conservation across the vdW gap.

**Summary**

In summary, we realized twisted magnetic tunnel junctions based on $La_{2/3}Sr_{1/3}MnO_3$ freestanding bilayers prepared by epitaxial lift-off technique. The clear observation of tunnelling magnetoresistance and tunnelling anisotropic magnetoresistance in the twisted junction reveals conservation of spin and in-plane momentum degrees of freedom across the vdW junction. Although tunnelling physics in twisted devices has been limited to two-dimensional materials so far[5], our demonstration will extend such explorations to a wide variety of three-dimensional transition metal oxides showing strongly correlated physical properties such as high-temperature superconductors[44,45].

## References

1. Geim, A. K. & Grigorieva, I. V. Van der Waals heterostructures. Nature **499**, 419–425 (2013).
2. Liu, Y., Weiss, N. O., Duan, X., Cheng, H. C., Huang, Y. & Duan, X. Van der Waals heterostructures and devices. Nat. Rev. Mat. **1**, 16042 (2016).
3. Suárez Morell, E., Correa, J. D.., Vargas, P., Pacheco, M. & Barticevic Z. Flat bands in slightly twisted bilayer graphene: Tight-binding calculations. Phys. Rev. B **82**, 121407 (2010).
4. Bistritzer R. & MacDonald, A. H. Moiré bands in twisted double-layer graphene. Proc. Natl. Acad. Sci. U.S.A. **108**, 12233–12237 (2011).
5. Mishchenko, A. et al. Twist-controlled resonant tunnelling in graphene/boron nitride/graphene heterostsructures. Nat. Nanotech. **9**, 808 (2014).
6. Cao, Y., Fatemi, V., Fang, S., Watanabe, K., Taniguchi, T., Kaxiras, E. & Jarillo-Herrero, P. Unconventional superconductivity in magic-angle graphene superlattices. Nature **566**, 43–50 (2018).
7. Sharpe, A. L., Fox, E. J., Barnard, A. W., Finney, J., Watanabe, K., Taniguchi, T., Kastner, M. A. & Goldhaber-Gordon, D. Emergent ferromagnetism near three-quarters filling in twisted bilayer graphene. Science **365**, 605–608 (2019).

8. Zheng, Z. et al. Unconventional ferroelectricity in moiré heterostructures. Nature **588**, 71–76 (2020).

9. Li, T. et al. Quantum anomalous Hall effect from intertwined moiré bands. Nature **600**, 641–646 (2021).

10. Imada, M., Fujimori, A. & Tokura, Y. Metal-insulator transitions. Rev. Mod. Phys. **70** 1039 (1998).

11. Hwang, H. Y. et al. Emergent phenomena at oxide interfaces. Nat. Mater. **11**, 103–113 (2012).

12. Mannhart, J. & Schlom, D. G. Oxide interfaces-an opportunity for electronics. Science 327, 1607–1611 (2010).

13. Chen, X. et al. Moiré engineering of electronic phenomena in correlated oxides. Nature Phys. **16**, 631-635 (2020).

14. Lu, D., Baek, D. J., Hong, S. S., Kourkoutis, L. F., Hikita, Y. & Hwang, H. Y. Synthesis of freestanding single-crystal perovskite films and heterostructures by etching of sacrificial water-soluble layers. Nat. Mat. **15**, 1255–1260 (2016).

15. Pesquera, D. et al., Beyond Substrates: Strain Engineering of Ferroelectric Membranes. Adv. Mater. **32**, 2003780 (2020).

16. Atsumi, R., Shiogai, J., Yamazaki, T., Seki, T., Ueda, K. & Matsuno, J. Impact of

epitaxial strain relaxation on ferromagnetism in a freestanding $La_{2/3}Sr_{1/3}MnO_3$ membrane. Jpn. J. Appl. Phys. **62**, 100902 (2023).

17. Wu, P. C., Twisted oxide lateral homostructures with conjunction tunability. Nature Commun. **13**, 2565 (2022).

18. Shen, J., Dong, Z., Qi, M., Zhang, Y., Zhu, C., Wu, Z., & Li, D. Observation of Moiré Patterns in Twisted Stacks of Bilayer Perovskite Oxide Nanomembranes with Various Lattice Symmetries. ACS Appl. Mater. Interfaces **14**, 50386–50392 (2022).

19. Li, Y. et al. Stacking and twisting of freestanding complex oxide thin films. Adv. Mater. **34**, 2203187 (2022).

20. Pryds, N., Park, D. S. Jespersen, T. S., Yun, S. Twisted oxide membranes: A perspective. APL Mater. **12**, 010901 (2024).

21. Sánchez-Santolino, G. et al. A 2D ferroelectric vortex pattern in twisted $BaTiO_3$ freestanding layers. Nature **626**, 529-534 (2024).

22. Sha, H., Zhang, Y., Ma, Y., Li, W., Yang, W., Cui, J., Li, Q., Huang, H. & Yu., R. Polar vortex hidden in twisted bilayers of paraelectric $SrTiO_3$. Nat. Commun. **15**, 10915 (2024).

23. Shahed, N. A., Samanta, K., Elekhtiar, M., Huang, K., Eom, C. B., Rzchowski, M. S., Belashchenko, K. D., & Tsymbal, E. Y. Prediction of polarization vortices, charge modulation, flat bands, and moiré magnetism in twisted oxide bilayers. Phys. Rev. B **111**, 195420 (2025).

24. Kim, M. S. et al. Charge Disproportionation at Twisted $SrTiO_3$ Bilayer Interface Driven by Local Atomic Registry. ASC Nano **19,** 39714-39724 (2025).
25. Zhang, B., Yun C., & MacManus-Driscoll, J. L., High Yield Transfer of Clean Large-Area Epitaxial Oxide Thin Films. Nano-Micro Lett. **13**, 39 (2021).
26. Li, B., Liu, Y., Liu, J., See, J. S. Y., Xing, K., Qi, D. C., & Wang, X. R., Emergent Freestanding Complex Oxide Membranes for Multifunctional Applications. Adv. Mater. **38**, e21388 (2026).
27. Krogh Andersen, I. G., Krogh Andersen, E., Norby, P. & Skou, E. Determination of Stoichiometry in Lanthanum Strontium Manganates(III)(IV) by Wet Chemical Methods. J. Solid State Chem. **113**, 320–326 (1994).
28. Arai, T. et al. Reduced dead layers and magnetic anisotropy change in $La_{2/3}Sr_{1/3}MnO_3$ membranes released from an $SrTiO_3$ substrate. Appl. Phys. Lett. **124**, 062403 (2024).
29. Simmons, J. G. Generalized formula for the electric tunnel effect between similar electrodes separated by a thin insulating film. J. Appl. Phys. **34**, 1793–1803 (1963).
30. Weston, A. et al., Atomic reconstruction in twisted bilayers of transition metal dichalcogenides. Nat. Nanotech. **15**, 592–597 (2020).

31. Zhang, Y. et al., "Tear-And-Stack" Twisted $SrTiO_3$ Moiré Superlattices for Precise Interfacial Reconstruction and Polar Topology. Adv. Mater. **38**, e19300 (2026).

32. Miyazaki, T. & Tezuka, N. Giant magnetic tunneling effect in $Fe/Al_2O_3/Fe$ junction. J. Magn. Magn. Mater. **139**, L231–L234 (1995).

33. Julliere, M. Tunneling between ferromagnetic films. Phys. Lett. **54A**, 225–226 (1975).

34. Bowen, M. et al. Spin-Polarized Tunneling Spectroscopy in Tunnel Junctions with Half-Metallic Electrodes. Phys. Rev. Lett. **95**, 137203 (2005).

35. Obata, R. et al. Pseudotunnel Magnetoresistance in Twisted van der Waals $Fe_3GeTe_2$ Homojunctions. Adv. Mater. **37**, 2411459 (2024).

36. Gould, C. et al. Tunneling Anisotropic Magnetoresistance: A Spin-Valve-Like Tunnel Magnetoresistance Using a Single Magnetic Layer. Phys. Rev. Lett. **93**, 117203 (2004).

37. Ciorga, M. et al. TAMR effect in (Ga,Mn)As-based tunnel structures. New J. Phys. **9**, 351 (2007).

38. Moser, J., Matos-Abiague, A., Schuh, D., Wegscheider, W., Fabian, J. & Weiss, D., Tunneling anisotropic magnetoresistance and spin-orbit coupling in Fe/GaAs/Au tunnel junctions. Phys. Rev. Lett. **99**, 056601 (2007).

39. Matos-Abiague, A. & Fabian, J. Anisotropic tunneling magnetoresistance and tunneling anisotropic magnetoresistance: Spin-orbit coupling in magnetic tunnel junctions. Phys. Rev. B **79**, 155303 (2009).

40. Shiogai, J. et al. In-plane tunneling anisotropic magnetoresistance in (Ga,Mn)As/GaAs Esaki diodes in the regime of the excess current. Appl. Phys. Lett. **106**, 262402 (2015).

41. Burton, J. D. & Tsymbal, E. Y., Tunneling anisotropic magnetoresistance in a magnetic tunnel junction with half-metallic electrodes. Phys. Rev. B **93**, 024419 (2016).

42. López-Mir, L. et al. Anisotropic sensor and memory device with a ferromagnetic tunnel barrier as the only magnetic element. Sci. Rep. **8**, 861 (2018).

43. Anh, L. D., Okamoto, N., Seki, M., Tabata, H., Tanaka M., & Ohya, S. Hidden peculiar magnetic anisotropy at the interface in a ferromagnetic perovskite-oxide heterostructure. Sci. Rep. **7**, 8715 (2017).

44. Zhu, Y. et al. Presence of *s*-wave pairing in Josephson junctions made of twisted ultrathin $Bi_2Sr_2CaCu_2O_{8+x}$ Flakes. Phys. Rev. X **11**, 031011 (2021).

45. Zhao, S. Y. F. et al. Time-reversal symmetry breaking superconductivity between twisted cuprate superconductors. Science **382**, 1422–1427 (2023).

46. Gong, L., Wei, M., Yu, R., Ohta, H. & Katayama, T. Significant suppression of cracks in freestanding perovskite oxide flexible sheets using a capping oxide layer. ACS Nano **16** 21013–21019 (2022).

47. Momma, K. & Izumi, F. VESTA 3 for three-dimensional visualization of crystal, volumetric and morphology data. J. Appl. Crystallogr. **44** 1272 (2011).

## Materials and Methods

**Sample fabrication and structural characterization.** The $La_{2/3}Sr_{1/3}MnO_3$ membranes were prepared by epitaxial lift-off technique[16]. First, a thin-film heterostructure consisting of $AlO_x$ cap/80-nm-thick $La_{2/3}Sr_{1/3}MnO_3/Sr_3Al_2O_6$ sacrificial layer was grown on $SrTiO_3(001)$ substrate by pulsed-laser deposition. The $Sr_3Al_2O_6$ was grown at substrate temperature $T_{sub}$ = 850°C and oxygen pressure $P_{O2} = 1.0 \times 10^{-5}$ Torr. The $La_{2/3}Sr_{1/3}MnO_3$ was grown at $T_{sub}$ = 750°C and $P_{O2} = 1.0 \times 10^{-1}$ Torr. The $AlO_x$ cap layer, deposited at room temperature in vacuum, offers a protection layer for suppression of crack formation during the lift-off process[16,46] as well as for eliminating residual contaminant directly on the $La_{2/3}Sr_{1/3}MnO_3$ membrane. The structural characterization using x-ray diffraction (XRD) confirms epitaxial growth of $La_{2/3}Sr_{1/3}MnO_3$ and $Sr_3Al_2O_6$ thin films (see Supplementary Fig. **S1a**). Next, polydimethylsiloxane (PDMS) sheet was attached on the surface of the heterostructure as a supporter and the stack of PDMS/$AlO_x$/$La_{2/3}Sr_{1/3}MnO_3/Sr_3Al_2O_6$ on $SrTiO_3$ substrate was dipped into distilled water so that the $Sr_3Al_2O_6$ layer was dissolved and the $AlO_x$/$La_{2/3}Sr_{1/3}MnO_3$ membrane was lifted off. The persistence of the *c*-axis orientation of the $La_{2/3}Sr_{1/3}MnO_3$ layer in the released membrane was confirmed by XRD (see Supplementary Fig. **S1b**). The $AlO_x$/$La_{2/3}Sr_{1/3}MnO_3$ membrane was then transferred onto the thermally-oxidized silicon substrate and the top $AlO_x$ cap layer was removed by Ar ion milling, exposing the clean surface of the bottom $La_{2/3}Sr_{1/3}MnO_3$ membrane. Another $AlO_x/La_{2/3}Sr_{1/3}MnO_3$ membrane was stamped on the bottom $La_{2/3}Sr_{1/3}MnO_3$ membrane with target in-plane twisted angle $\theta$ to form the twisted $La_{2/3}Sr_{1/3}MnO_3/La_{2/3}Sr_{1/3}MnO_3$ bilayer junction. Then, Pt electrodes were formed by photolithography, Ar ion milling, sputtering, and lift-off technique. Finally, the junction samples were annealed in air to compensate for oxygen deficiency possibly formed during the milling process. The twisted junction interface was characterized by scanning electron microscopy (SEM), cross-sectional bright field (BF) transmission electron micrography (TEM), and high angle annular dark field (HAADF) scanning transmission electron micrography (STEM). The crystal structures were drawn by VESTA (Ref. 47).

**Electrical resistance and magnetization measurements.** The samples were mounted in either variable temperature insert equipped with a superconducting magnet (Oxford Instruments plc) or Physical Property Measurement System (PPMS, Quantum Design Inc.). *I-V* characteristic measurements were performed by DC method with pulsed current. MR measurements were carried out by DC method using source-measure unit and nanovoltmeter for high bias voltage or AC method using lock-in amplifier for low bias voltage. The magnetization properties of reference un-patterned membrane samples on silicon substrate were characterized using Magnetic Property Measurement System (MPMS, Quantum Design Inc.) (see Supplementary Section **3**).

**Figure 1 | Tunnelling across the twisted junction of $La_{2/3}Sr_{1/3}MnO_3$ membranes. a,** Top-view illustration of twisted junction of $La_{2/3}Sr_{1/3}MnO_3$. **b,** Schematic for magnetic tunnel junction of twisted $La_{2/3}Sr_{1/3}MnO_3$ with van der Waals (vdW) gap serving as a tunnel barrier. **c,** Conservation of spin and in-plane momentum degree of freedom of electron across the tunnel barrier. **d,** Schematic for device fabrication process. The $AlO_x$ cap/80-nm-thick $La_{2/3}Sr_{1/3}MnO_3/Sr_3Al_2O_6$ heterostructure is dipped into water to selectively remove the $Sr_3Al_2O_6$ layer. An exfoliated $AlO_x/La_{2/3}Sr_{1/3}MnO_3$ membrane is transferred on the thermally oxidized silicon substrate and $AlO_x$ is removed by Ar ion milling. Another $AlO_x/La_{2/3}Sr_{1/3}MnO_3$ membrane is stamped on it to form $La_{2/3}Sr_{1/3}MnO_3/La_{2/3}Sr_{1/3}MnO_3$ junction. Twist angle $\theta$ is defined as in-plane rotation angle of the top membrane anticlockwise with respect to the bottom membrane.

**Figure 2 | a,** Optical micrograph of exfoliated membrane on PDMS. **b,** Scanning electron micrograph of a typical $La_{2/3}Sr_{1/3}MnO_3$ junctions. **c,d** Cross-sectional (**c**) bright field (BF) transmission electron microscope (TEM) and (**d**) high angle annular dark field (HAADF) scanning transmission electron microscope (STEM) images of 45°-junction. Insets in **d** show electron diffraction patterns acquired separately from the top and bottom membranes. **e**, Temperature dependences of two-terminal resistance $R_J$ across the

$La_{2/3}Sr_{1/3}MnO_3$/$La_{2/3}Sr_{1/3}MnO_3$ junction (red solid line, $R_J$) and three-terminal resistances across the Pt/$La_{2/3}Sr_{1/3}MnO_3$ junction for top and bottom membranes (green, $R_{3T}^{top}$ and blue $R_{3T}^{btm}$) for 45°-twisted junction. **f,** *I*-*V* characteristics for 45°-junction at $T$ = 4.2 K Dashed line is fitting curve based on Ref. 29. **g,** Temperature dependences of a product of $R_J$ and junction area $A$ for $\theta \sim 0°$ (green), 45° (red), and 70° (blue).

**Figure 3 | Observation of tunnelling magnetoresistance (TMR) effect in twisted junction of $La_{2/3}Sr_{1/3}MnO_3$ membranes. a,** Magnetoresistance (MR) in $R_J(B)$ for the twisted junctions of $La_{2/3}Sr_{1/3}MnO_3$ membranes with the twist angle $\theta \sim 45$ and 0° at $V_b$ ~ 40 mV and that in $R_{3T}^{btm}(B)$ for 45°-junction. **b,** MR in $R_J(B)$ for 45°-junction when applied bias voltage $V_b$ = 2 (top), 4, 30, and 200 mV (bottom). **c,** MR ratio for 0°- (green circles) and 45°-junctions (red circles), defined as a vertical arrow in **b**, as a function of $V_b$. All MR measurements were performed at $T$ = 2 K. **d,** Full loop (black and red lines) and minor loop (blue line) of MR in $R_J(B)$ of 0°-junction. The parallel configuration between magnetizations of top and bottom membranes (red circle) was obtained by sweeping $B$ from +1 T to 0 T while its anti-parallel configuration (blue circle) was obtained by sweeping B from +1 T to −16.5 mT and back to 0 T. **e,** $R_J(T)$ curves obtained by increasing $T$ starting from the parallel configuration [$R_P(T)$, red] and anti-parallel

configuration [$R_{AP}(T)$, blue] at $B$ = 0 T. (Inset) Temperature dependence of MR ratio defined by $(R_{AP} - R_P)/R_P$.

**Figure 4 | In-plane tunnelling anisotropic magnetoresistance (TAMR) effect. a,** $R_J(B)$ for 70°-junction when in-plane magnetic field $B$ is applied along [100] of the bottom membrane. **b,** $R_J(B)$ for 0°-junction when $B$ is applied along $\phi_B$ = 170° (blue lines) and 74° (red lines) from [100] of the bottom membrane. The solid and dashed lines correspond to the data obtained for down and up sweep, respectively. **c,** $\phi_B$ dependence of $R_J$ at $B$ = 1 T, normalized by the value when $\boldsymbol{B}$ // [110] for 0°-junction. **d,** Rhombohedral distortion of the $La_{2/3}Sr_{1/3}MnO_3$. Orange lines represent rhombohedral unit-cell. **e,** Two-dimensional schematic of the corner shared $MnO_6$ network on pseudocubic (001) plane.

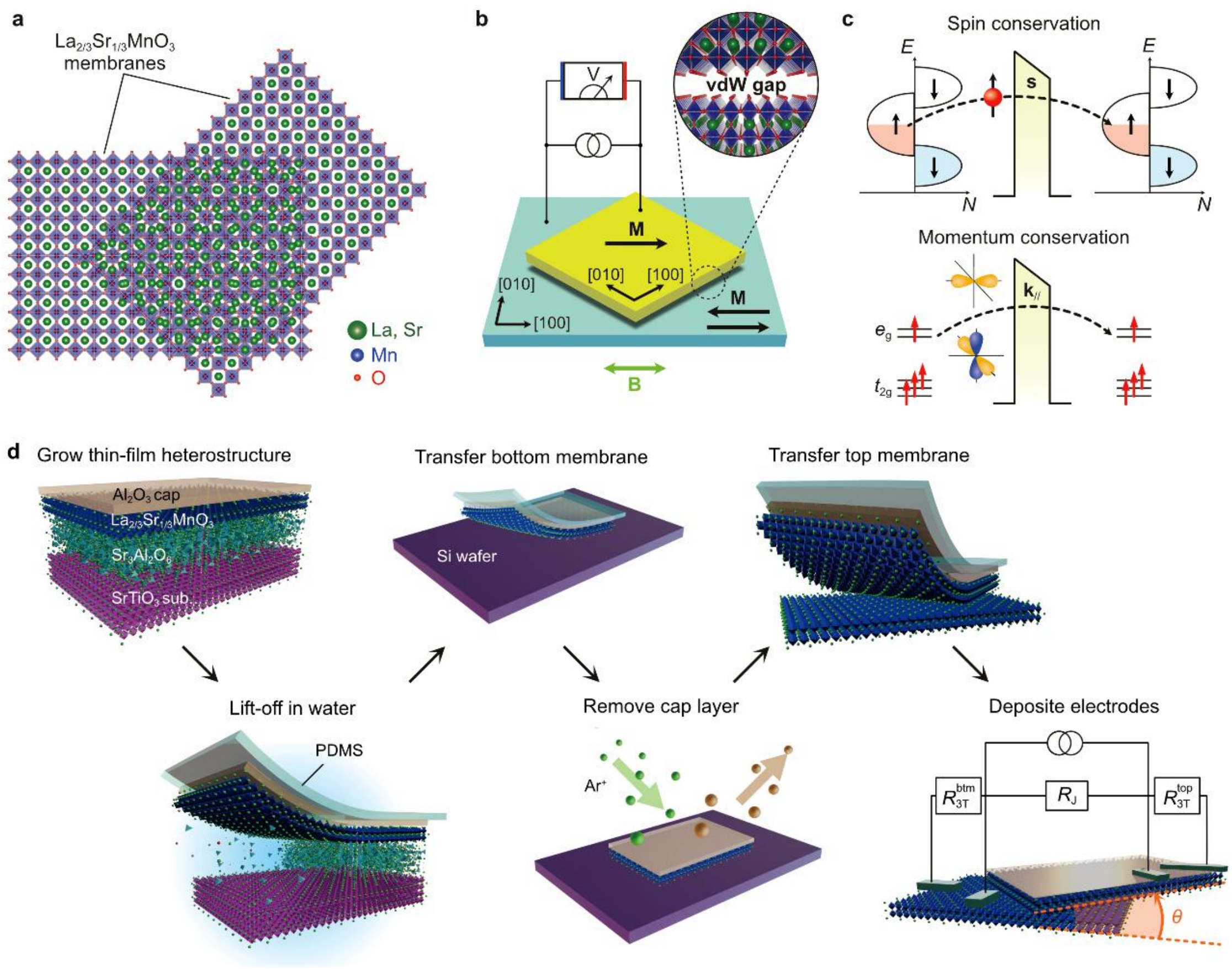

Figure 1 Kanda *et al*

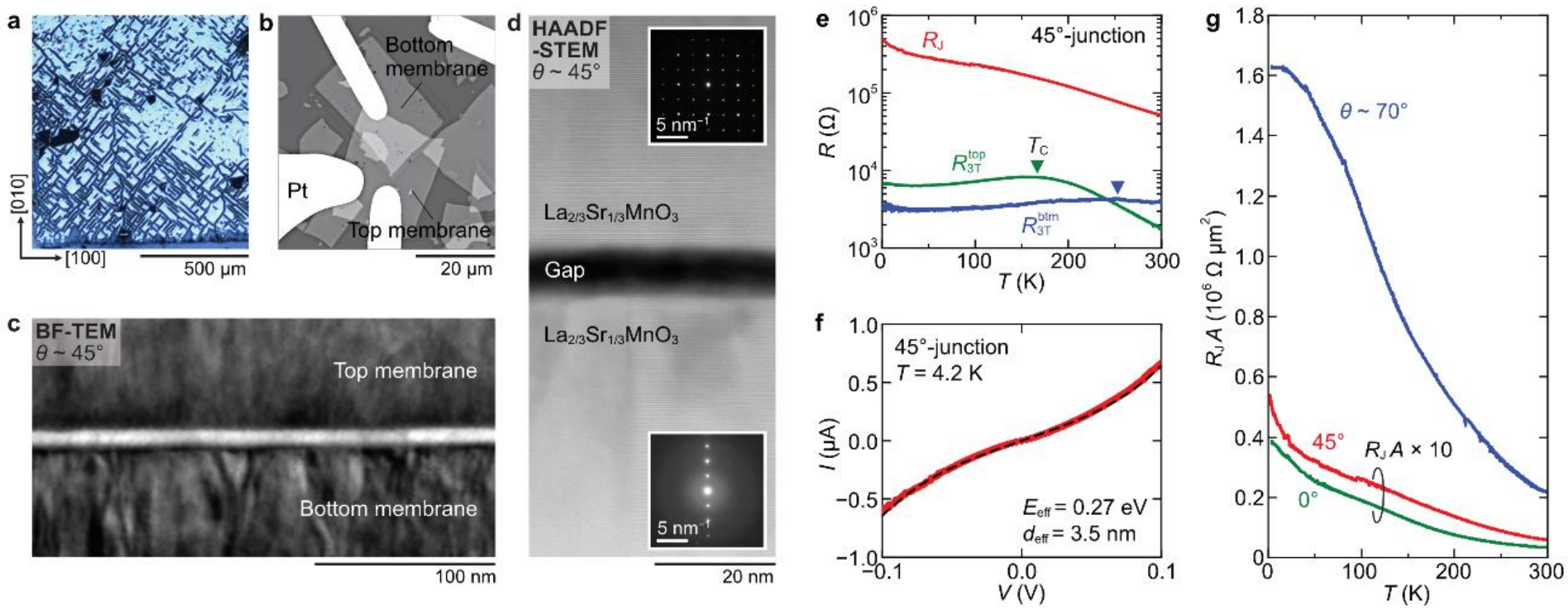

Figure 2 Kanda *et al*

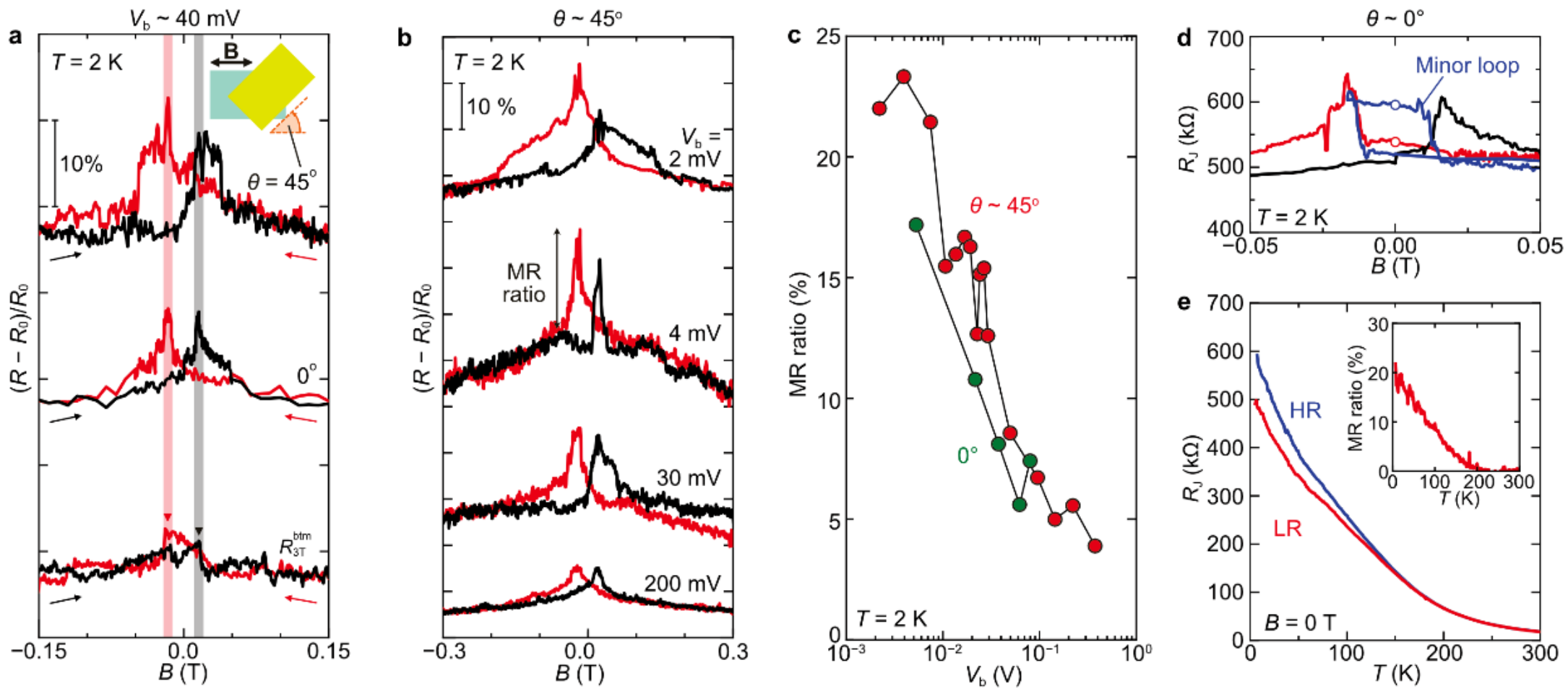

Figure 3 Kanda *et al*

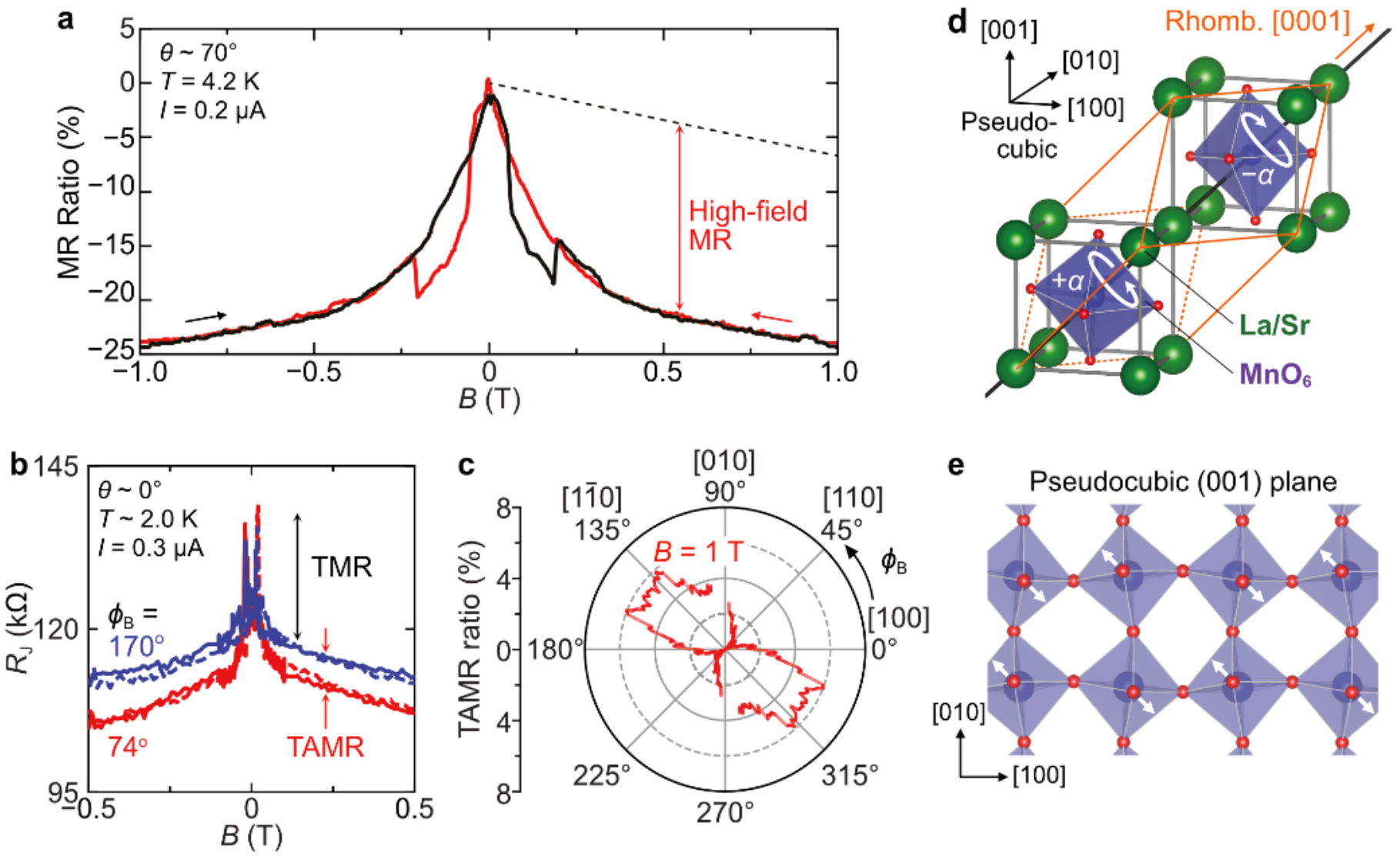

Figure 4 Kanda *et al*

Supplementary Information for

# Magnetic tunnel junctions of twisted $La_{2/3}Sr_{1/3}MnO_3$ freestanding bilayers

Kota Kanda[1], Shiho Irie[1], Kodai Niitsu[2], Masayuki Ochi[1,3], Takumi Yamazaki[4], Kouta Kondou[5], Kohei Ueda[1,5], Takeshi Seki[4,6,7], Jobu Matsuno[1,5], and Junichi Shiogai[1,5,†]

[1]*Department of Physics, The University of Osaka, Toyonaka, Osaka 560-0043, Japan*

[2]*Center for Basic Research on Materials, National Institute for Materials Science (NIMS), Tsukuba, Ibaraki 305-0047, Japan*

[3]*Forefront Research Center, The University of Osaka, Toyonaka, Osaka 560-0043, Japan*

[4]*Institute for Materials Research, Tohoku University, Sendai, Miyagi 980-8577, Japan*

[5]*Division of Spintronics Research Network, Institute for Open and Transdisciplinary Research Initiatives, The University of Osaka, Suita, Osaka 565-0871, Japan*

[6]*Center for Science and Innovation in Spintronics, Tohoku University, Sendai, Miyagi 980-8577, Japan*

[7]*International Center for Synchrotron Radiation Innovation Smart, Tohoku University, Sendai, Miyagi 980-8577, Japan*

## 1. Sample fabrication and structural characterization

Thin-film heterostructures consisting of $AlO_x$ cap/80-nm-thick $La_{2/3}Sr_{1/3}MnO_3/Sr_3Al_2O_6$ sacrificial layer were prepared on $SrTiO_3(001)$ substrate by pulsed-laser deposition with a laser wavelength of 248 nm (Ref. S1). The $Sr_3Al_2O_6$ layer was grown at substrate temperature $T_{sub}$ = 850°C and oxygen pressure $P_{O2} = 1.0 \times 10^{-5}$ Torr. The $La_{2/3}Sr_{1/3}MnO_3$ layer was grown at $T_{sub}$ = 750°C and $P_{O2} = 1.0 \times 10^{-1}$ Torr. The $AlO_x$ cap layer was deposited at room temperature in vacuum. Figure **S1a** and **S1b** show x-ray diffraction (XRD) patterns for a heterostructure of $AlO_x/La_{2/3}Sr_{1/3}MnO_3/Sr_3Al_2O_6$ and an exfoliated membrane of $AlO_x/La_{2/3}Sr_{1/3}MnO_3$, indicating that sacrificial layer of $Sr_3Al_2O_6$ was completely removed in the lift-off process and the single-crystalline $La_{2/3}Sr_{1/3}MnO_3$ membrane is exfoliated from the substrate. Figure **S1c** shows top-view optical micrograph of the exfoliated $La_{2/3}Sr_{1/3}MnO_3$ membrane. Many cracks are observed especially along <100> and <110> direction of $La_{2/3}Sr_{1/3}MnO_3$, which we assigned to twin boundaries due to rhombohedral distortion.

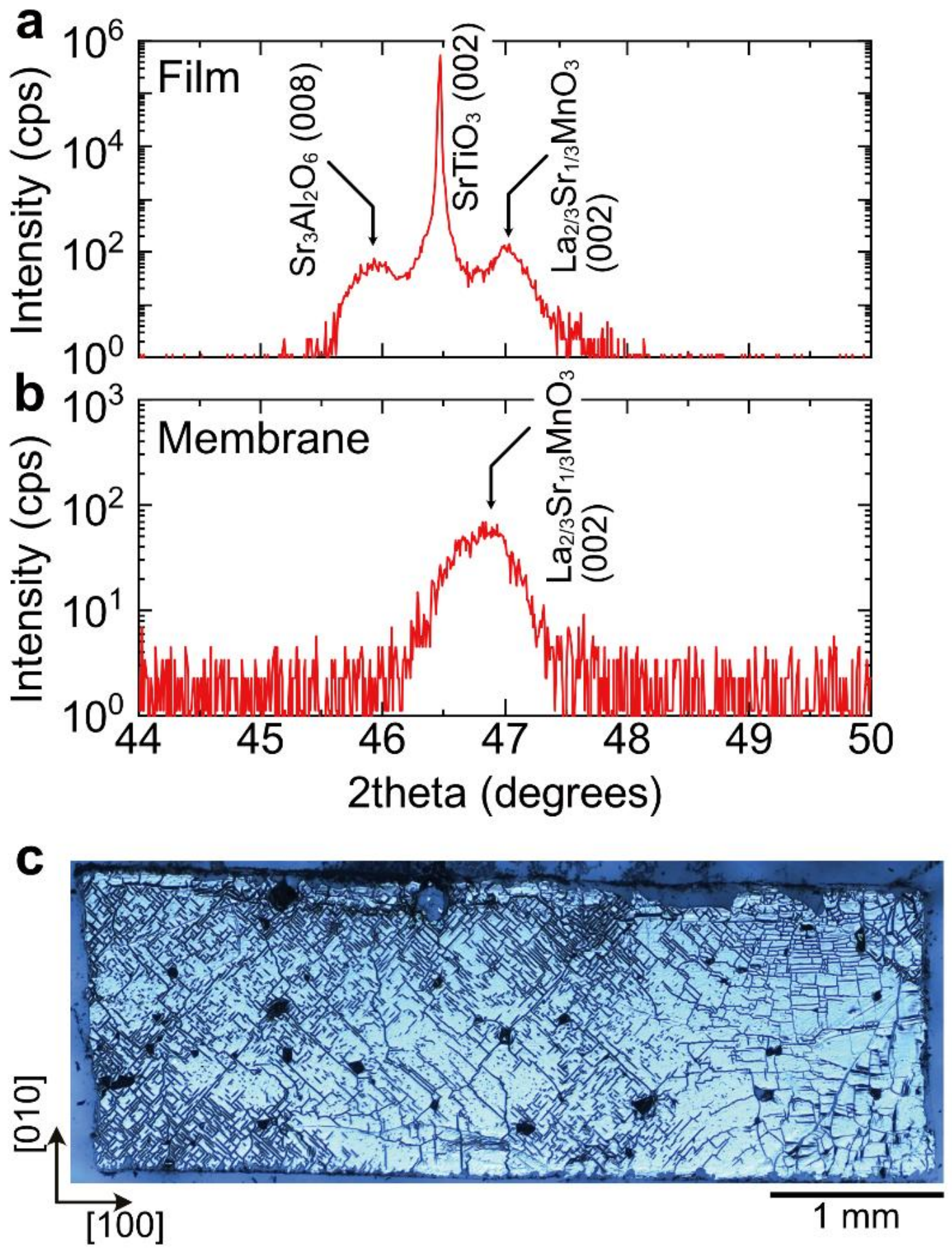

**Figure S1 | Structural analysis. a,b** X-ray diffraction patterns of **a,** $AlO_x/La_{2/3}Sr_{1/3}MnO_3/Sr_3Al_2O_6$ thin-film heterostructure and **b,** exfoliated $AlO_x/La_{2/3}Sr_{1/3}MnO_3$ membrane. **c,** Optical micrograph of exfoliated $La_{2/3}Sr_{1/3}MnO_3$ membrane.

## 2. Three-terminal resistance measurements of Pt/$La_{2/3}Sr_{1/3}MnO_3$ junctions

Figures **S2a** and **S2b** show a measurement configuration with a cross-sectional schematic of Pt/$La_{2/3}Sr_{1/3}MnO_3$/$La_{2/3}Sr_{1/3}MnO_3$/Pt junction, and corresponding equivalent circuit. In our experiments, the spin-valve effect is measured by magnetic-field ($B$) dependence two-terminal resistance $R_J$, which should consist of three resistive contributions at (I) $La_{2/3}Sr_{1/3}MnO_3$/$La_{2/3}Sr_{1/3}MnO_3$ junction, denoted as $R_I$, (II) $La_{2/3}Sr_{1/3}MnO_3$ single membrane, $R_{II}$, and (III) Pt/$La_{2/3}Sr_{1/3}MnO_3$ junction, $R_{III}$. We first exclude $B$ dependence of (II) such as the anisotropic magnetoresistance (AMR) because MR ratio measured in a single membrane is much smaller than that in $R_J$ and $R_{II}$ value of a few kΩ is three-order of magnitude smaller than $R_J$.

Only the contribution (III), corresponding to $R_{III}I$, is measured by the three-terminal voltages $V_{3T}^{top}$ and $V_{3T}^{btm}$. Figures **S2c-S2e** show the magnetic-field dependence resistances $R_J$, and concomitantly measured $R_{3T}^{top} = V_{3T}^{top}/I$ and $R_{3T}^{btm} = V_{3T}^{btm}/I$ at various current $I$. While $R_J(B)$ exhibits spin-valve-like magnetoresistance, $R_{3T}^{top}(B)$ and $R_{3T}^{btm}(B)$ do not show such spin-valve-like magnetoresistance. Therefore, we conclude that the spin-valve signal in $R_J$ originates from the $La_{2/3}Sr_{1/3}MnO_3$/$La_{2/3}Sr_{1/3}MnO_3$ junction.

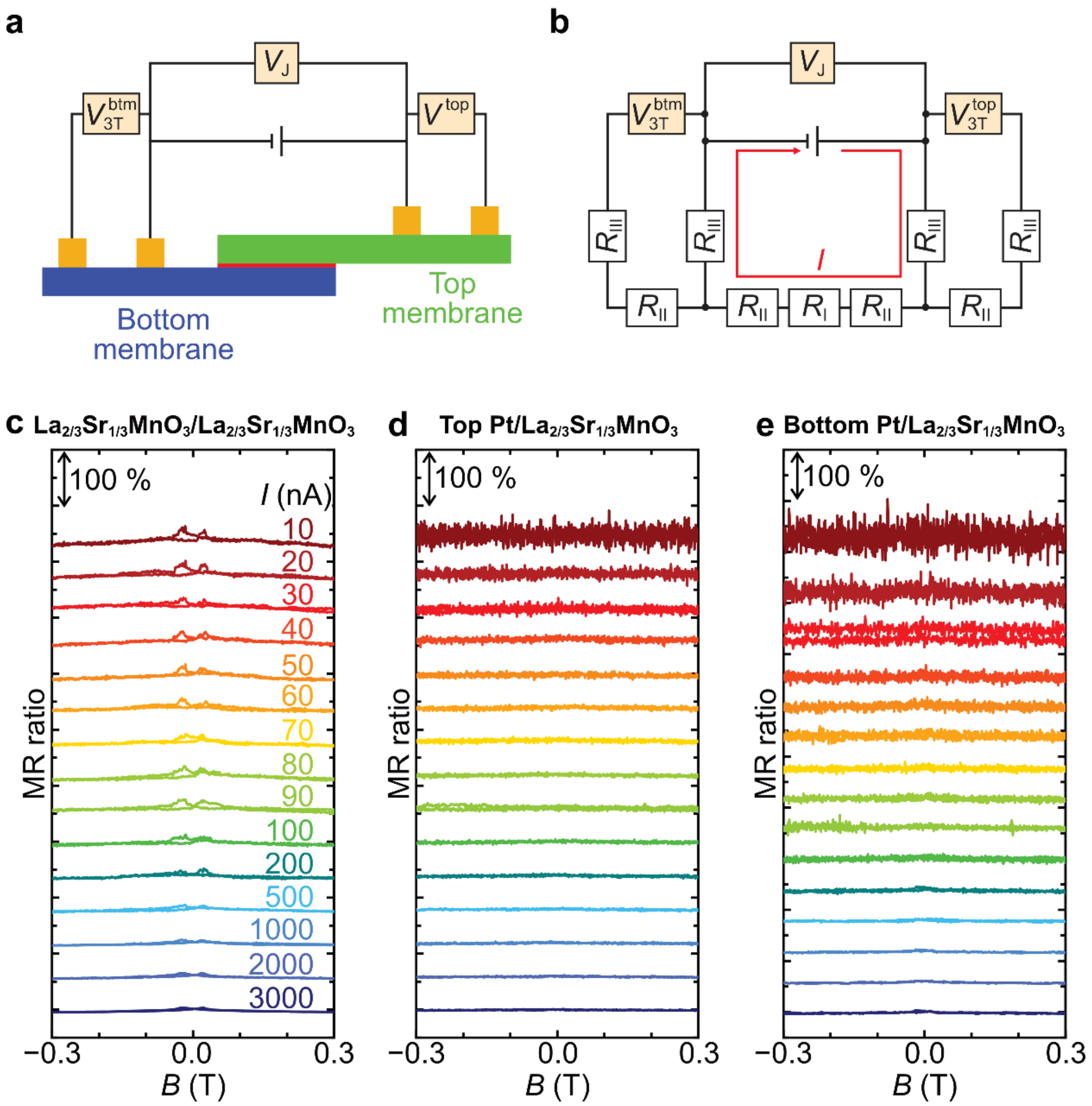

**Figure S2 | Magnetoresistance across the $La_{2/3}Sr_{1/3}MnO_3/La_{2/3}Sr_{1/3}MnO_3$ and $Pt/La_{2/3}Sr_{1/3}MnO_3$ Junction. a,** Two-terminal and three-terminal measurement configurations and **b,** equivalent circuits taking $R_I$, $R_{II}$, and $R_{III}$ contributions into account. Magnetic field dependence of **c,** $R_J(B)$, **d,** $R_{3T}^{top}(B)$, and **e,** $R_{3T}^{btm}(B)$ for 45°-twisted junctions at various applied currents.

## 3. Magnetization properties of $La_{2/3}Sr_{1/3}MnO_3$ membrane

Magnetic properties of $La_{2/3}Sr_{1/3}MnO_3$ membrane were evaluated by magnetization measurements. Figure **S3a** shows magnetic hysteresis loop for the same $La_{2/3}Sr_{1/3}MnO_3$ membrane, measured after each process of transferring on thermally oxidized Si substrate (black), Ar ion milling (blue), and annealing in air (red). Figure **S3b** shows temperature dependence of magnetization under the magnetic field of 0.15 T along [100] of $La_{2/3}Sr_{1/3}MnO_3$. The $T_C$ was evaluated by a kink in temperature derivative of magnetization (bottom panel). The milling and annealing condition are identical to those used for fabricating the twisted junctions. After the series of fabrication processes, while the coercive field is slightly reduced from 16 to 11 mT, the $T_C$ is unchanged at 334 K.

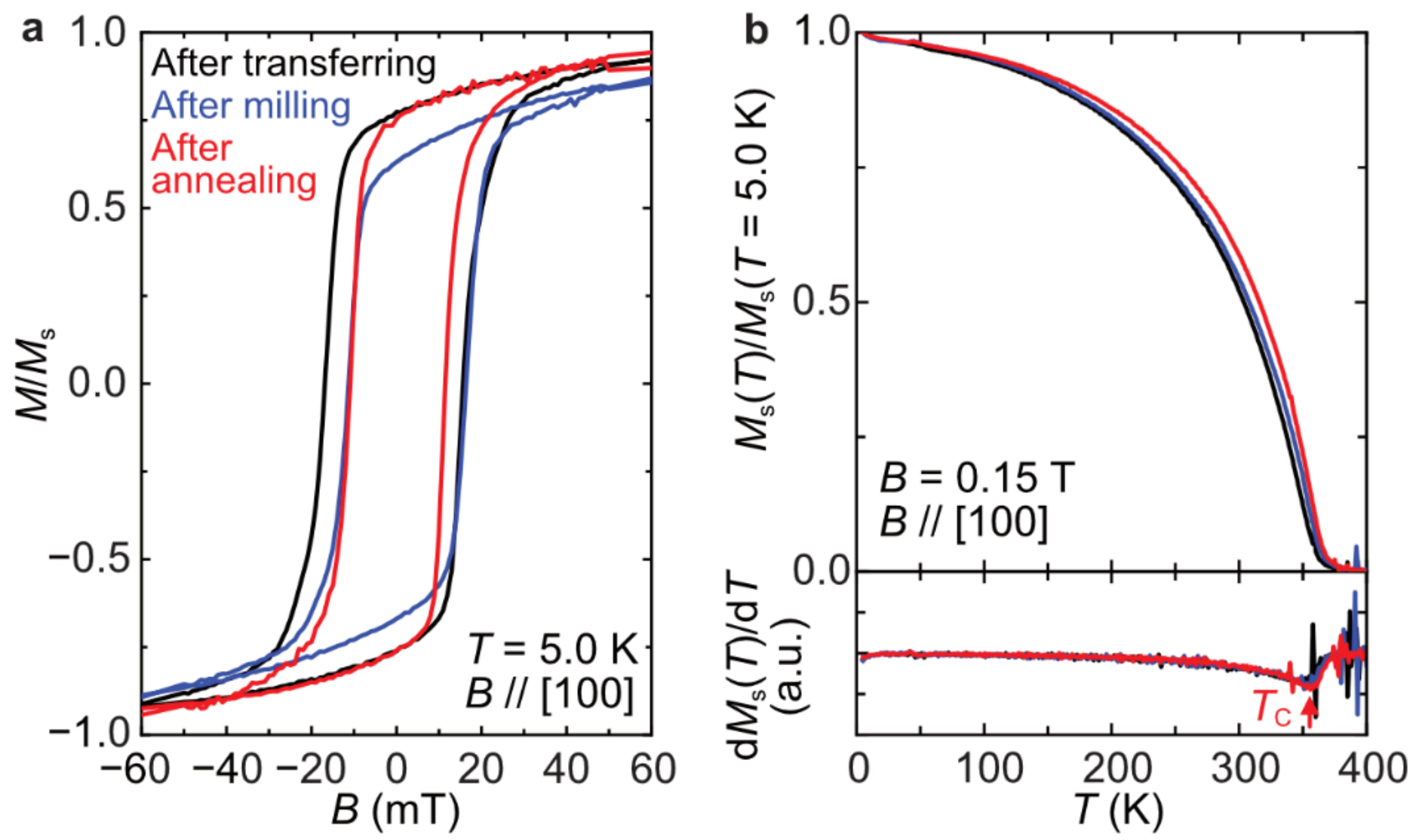

**Figure S3 | Magnetization properties. a,** Magnetic hysteresis loop and **b,** temperature dependence of magnetization of $La_{2/3}Sr_{1/3}MnO_3$ membrane measured after transferring (black), milling (blue), and annealing (red).

## 4. Tunnelling conductance properties across the twisted junctions

Figure **S4a** and **S4b** show *I*-*V* characteristics (red circles) measured at $T = 4.2$ K for 0°- and 70°-junctions. Overall, the upturn non-linear feature in the *I*-*V* curve is observed, indicating that tunnelling transport dominates the current. The *I*-*V* characteristics are fitted based on Simmons' formula[S2] (black dashed lines),

$$J = J_0 \left[ \bar{\phi} \exp\left(-A\sqrt{\bar{\phi}}\right) - (\bar{\phi} + eV) \exp\left(-A\sqrt{\bar{\phi} + eV}\right) \right], \qquad \text{(S1)}$$

$$\bar{\phi} = \frac{1}{\Delta s} \int_0^d \phi(x)\, \mathrm{d}x, \qquad \text{(S2)}$$

$$J_0 = \frac{e}{2\pi h (\Delta s)^2}, \qquad \text{(S3)}$$

$$A = \frac{4\pi \Delta s}{h} \sqrt{2m}. \qquad \text{(S4)}$$

Here, $e$ and $m$ are electron charge and mass, $h$ is Planck constant, $d$ is physical barrier thickness, and $V$ is applied voltage. Mean barrier height $\bar{\phi}$ and barrier width $\Delta s$ are fitting parameters, which we refer to $E_{\mathrm{eff}}$ and $d_{\mathrm{eff}}$ in the main text, respectively. The extracted values of $E_{\mathrm{eff}}$ and $d_{\mathrm{eff}}$ for the 0°-junction are 0.22 eV and around 3.6 nm, and those for the 70°-junction are 0.27 eV and 4.1 nm, which are comparable to those observed for the 45°-junction (main text).

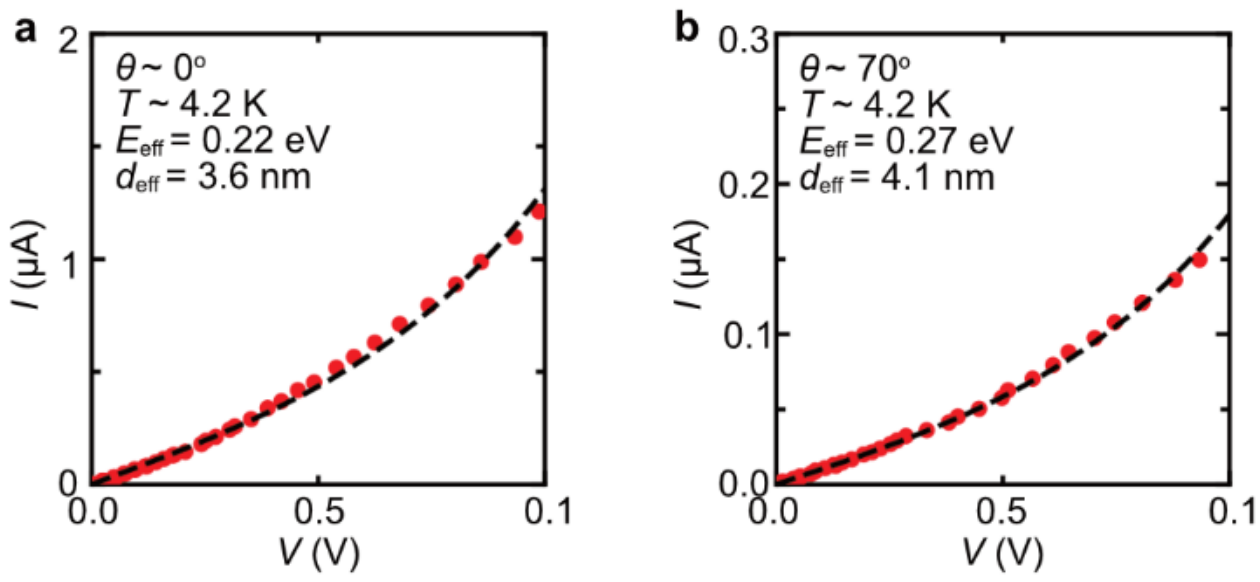

**Figure S4 | *I-V* characteristics for twisted junctions. a,b,** *I-V* characteristics with fitting results for **a,** 0°-junction and **b,** 70°-junction.

## 5. Electronic structure calculation for bulk $La_{2/3}Sr_{1/3}MnO_3$

Using density functional theory (DFT) calculations, we investigate bulk electronic band structure of $La_{2/3}Sr_{1/3}MnO_3$ at various orientations of magnetization vector **M**. To model the anisotropy in $La_{2/3}Sr_{1/3}MnO_3$ with the rhombohedral distortion, we took a 2×2×2 pseudocubic supercell containing eight Mn atoms with the same octahedral rotation as shown in Fig. **4**. Our calculations were performed using the generalized gradient approximation with the Perdew-Burke-Ernzerhof parameterization[S3] as implemented in Vienna *Ab initio* Simulation Package[S4-S8]. A fractional occupation of La and Sr was handled with virtual crystal approximation. First, we fixed the cubic lattice constant $a$ = 7.72 Å and optimized the internal coordinates until the Hellmann-Feynman force on each atom becomes smaller than 0.005 eV/Å. For this purpose, we used a plane-wave cutoff energy of 600 eV and a 6×6×6 $\boldsymbol{k}$ mesh without taking spin-orbit interaction into account. Ferromagnetism of Mn spins was assumed throughout calculations. After that, we performed self-consistent-field calculations and subsequent band calculations with three orientations of magnetization vector **M**, [100], [110], and [$1\bar{1}0$]. For this purpose, we used a plane-wave cutoff energy of 700 eV and a 7×7×7 $\boldsymbol{k}$ mesh with including spin-orbit interaction. Band $\boldsymbol{k}$ points were taken as X (π/a, 0, 0), Z (0, 0, π/a), and R (π/a, 0, π/a).

Figures **S5a** and **S5b** show calculated electronic band structures $E(\boldsymbol{k})$ for the cases when **M** is oriented along [110] and [$1\bar{1}0$], respectively, with a color weight representing an energy difference relative to the case of **M** // [100]. Note that significant anisotropy is not resolved in the density of state obtained by integrating $E(\boldsymbol{k})$ over $\boldsymbol{k}$-space for **M**//[110] and [$1\bar{1}0$]. Instead, we discuss $\boldsymbol{k}$-resolved energy difference between the cases of **M**// [110] and [$1\bar{1}0$]. The sign change of energy difference around the Γ-point is clearly observed when **M** is switched from [110] to [$1\bar{1}0$]. Although the maximum value of the energy difference along the Γ-X line is at most 1 meV for the electronic states having an energy within the range of −1 eV to +1 eV in our bulk calculation, it is suggested that the interfacial electric field pushes $t_{2g}$ bands toward Fermi energy $E_F$, enhancing the hybridization between $d_{z^2}$ and $d_{xz}/d_{yz}$ orbitals, according to the previous theoretical work considering $SrTiO_3/La_{2/3}Sr_{1/3}MnO_3$ interfaces[S9]. As a result, the anisotropy in transmission probability is expected to be largely enhanced around $E_F$. In our freestanding membrane, it can be speculated that such localized interface states can be created by, for example, surface oxide deficiency as reported previously[S10] and being consistent with the reduced Curie temperature at surface of the membrane (Fig. **2e**).

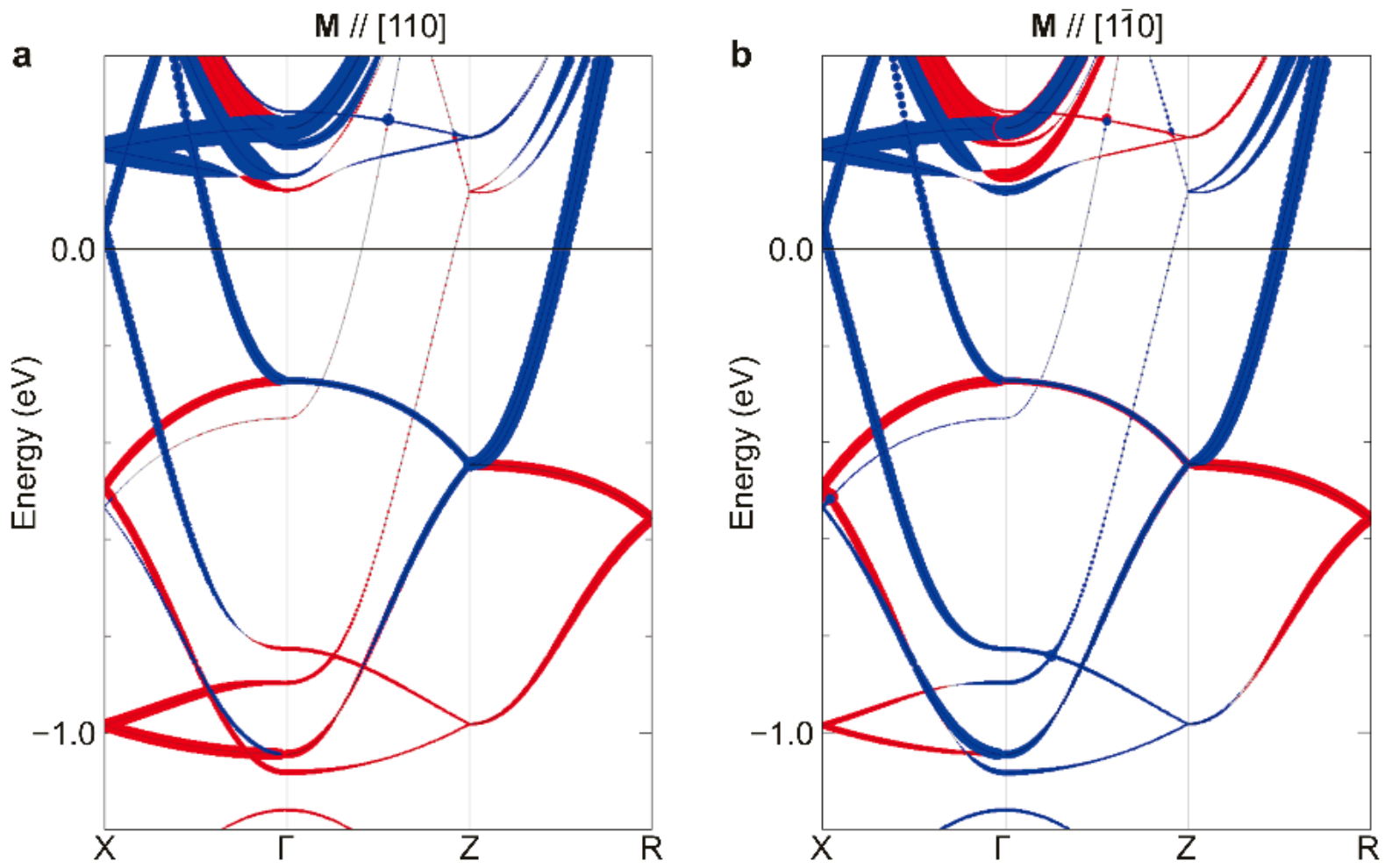

**Figure S5 | Electronics structure calculation.** Electronic band structures of ferromagnetic $La_{2/3}Sr_{1/3}MnO_3$ for the cases when **a,** **M** // [110] and **b,** **M** // [1$\bar{1}$0], respectively. Diameter of colored circles is proportional to the energy difference with respect to the case when **M** // [100]. Red and blue circles represent positive and negative signs, respectively.

**Supplementary References**

S1. Atsumi, R., Shiogai, J., Yamazaki, T., Seki, T., Ueda, K., & Matsuno, J. Impact of epitaxial strain relaxation on ferromagnetism in a freestanding $La_{2/3}Sr_{1/3}MnO_3$ membrane. Jpn. J. Appl. Phys. **62**, 100902 (2023).

S2. Simmons, J. G. Generalized formula for the electric tunnel effect between similar electrodes separated by a thin insulating film. J. Appl. Phys. **34**, 1793–1803 (1963).

S3. Perdew, J. P., Burke, K., and Ernzerhof, M., Generalized Gradient Approximation Made Simple. Phys. Rev. Lett. **77**, 3865 (1996).

S4. Kresse, G. & Hafner, J., *Ab initio* molecular dynamics for liquid metals. Phys. Rev. B **47**, 558 (1993).

S5. Kresse, G. & Hafner, J., *Ab initio* molecular-dynamics simulation of the liquid-metal–amorphous-semiconductor transition in germanium. Phys. Rev. B **49**, 14251 (1994).

S6. Kresse, G. & Furthmüller, J., Efficiency of ab-initio total energy calculations for metals and semiconductors using a plane-wave basis set. Comput. Mater. Sci. **6**, 15 (1996).

S7. Kresse, G. & Furthmüller, J., Efficient iterative schemes for *ab initio* total-energy calculations using a plane-wave basis set. Phys. Rev. B **54**, 11169 (1996).

S8. Kresse, G. & Joubert, D., From ultrasoft pseudopotentials to the projector augmented-wave method. Phys. Rev. B **59**, 1758 (1999).

S9. Burton, J. D. & Tsymbal, E. Y., Tunneling anisotropic magnetoresistance in a magnetic tunnel junction with half-metallic electrodes. Phys. Rev. B **93**, 024419 (2016).

S10. Arai, T. *et al*. Reduced dead layers and magnetic anisotropy change in $La_{2/3}Sr_{1/3}MnO_3$ membranes released from an $SrTiO_3$ substrate. Appl. Phys. Lett. **124**, 062403 (2024).